\documentclass[a4paper,12pt]{article}
\usepackage{epsfig}
\usepackage{graphicx}
\usepackage{dcolumn}
\usepackage{bm}
\usepackage{longtable}
\usepackage{amssymb}
\usepackage{amsfonts}
\usepackage[margin=1.0in]{geometry}
\usepackage{amsmath}


\usepackage{slashed}

\def\IC{\relax\,\hbox{$\inbar\kern-.3em{\rm C}$}}

\def\cL{\,\mathcal{L}}
\def\cM{\,\mathcal{M}}

\def\ii{ {\mathrm{i}} }
\newcommand{\eq}[1]{(\ref{#1})}

\DeclareFontFamily{U}{rsf}{}
\DeclareFontShape{U}{rsf}{m}{n}{
  <5> <6> rsfs5 <7> <8> <9> rsfs7 <10-> rsfs10}{}
\DeclareMathAlphabet\Scr{U}{rsf}{m}{n}

\makeatletter                             
\@addtoreset{equation}{section}                   
\makeatother                              
\usepackage{hyperref}

\begin{document}

\begin{center}
{\bf\LARGE N=1 and N=2 pure supergravities on a manifold with boundary} \\
\vskip 2 cm
{\bf  Laura Andrianopoli$^{1,2}$ and   Riccardo D'Auria$^1$ }
\vskip 8mm
 \end{center}
\noindent {\small{1. {\it  DISAT, Politecnico di Torino, Corso Duca
    degli Abruzzi 24, I-10129 Turin, Italy} \\
    2. {\it Istituto Nazionale di
    Fisica Nucleare (INFN) Sezione di Torino, Italy}
    \\
    \\
 email:    \texttt{laura.andrianopoli@polito.it}, \texttt{riccardo.dauria@polito.it}}
\vskip 1 cm

{\small
\begin{abstract}
Working in the geometric approach, we construct the lagrangians of $N=1$ and $N=2$ pure supergravity in four dimensions with negative cosmological constant, in the presence of a non trivial boundary of space-time.  We find that the supersymmetry invariance of the action requires the addition of topological terms which generalize at the supersymmetric level the Gauss--Bonnet term. Supersymmetry invariance is achieved without requiring Dirichlet boundary conditions on the fields at the boundary, rather we find that the boundary values of the fieldstrengths are dynamically fixed to  constant values in terms of the cosmological constant $\Lambda$. From a group-theoretical point of view this means in particular  the vanishing  of the $OSp(N|4)$-supercurvatures at the boundary.
\end{abstract}
}


 \section{Introduction}

 Gravity and supergravity lagrangians in the presence of a boundary have been studied in different contexts from the early seventies on.  The need of adding a boundary term to the gravity action, such as to implement Dirichlet boundary conditions for the metric field, was first pointed out in \cite{York:1972sj} and
 \cite{Gibbons:1976ue}, in  early attempts to study the quantization of gravity with a path integral approach, in order to have an action which depends only on the first derivatives of the metric.  More recently, the addition of boundary terms was considered in \cite{Horava:1996ma} to cancel gauge and gravitational anomalies in the Horava--Witten model in 11D.
Inclusion of boundary terms  is also an essential tool for the study of the AdS/CFT duality   \cite{Maldacena:1997re}, a well celebrated and far reaching  duality between string theory on asymptotically AdS space-time (times a compact manifold) and a quantum field theory living on the boundary.\footnote{The literature on the subject of AdS/CFT and on its developments in various directions as gauge/gravity correspondence is so huge that we limit to refer to the first publications and to general reviews containing  more extended reference lists.} The duality was tested very deeply at low energies, in the supergravity limit of string theory, where it implies a one-to-one correspondence between quantum operators $\mathcal O$ in the boundary conformal field theory and fields $\phi$ of the bulk supergravity theory and requires to supplement the supergravity action functional with appropriate boundary conditions $\phi^{(0)}$ for the supergravity fields, which act as sources for the operators of the CFT.
  As far as the metric field is concerned, in particular,  the bulk  metric is divergent near the boundary.  These divergences can be however disposed of successfully by the so called holographic renormalization \cite{holren} through the inclusion of appropriate counterterms at the boundary.

 The inclusion of boundary terms and counterterms to the bosonic sector of AdS supergravity has been extensively studied in many different contexts.   In particular, interesting results have been obtained in  \cite{Aros:1999id}, where it was shown that the addition of the topological Euler--Gauss-Bonnet term to the Einstein action of four dimensional AdS gravity leads to a background-independent definition of Noether charges,  without the need of imposing  Dirichlet boundary conditions on the fields.
  Such boundary term indeed regularizes the action and the related (background independent) conserved charges.

At the full supergravity level, boundary contributions  were considered from several authors, using different approaches, and in  particular in \cite{vanNieuwenhuizen:2005kg}, \cite{esposito}, \cite{Moss:2003bk}, \cite{Howe:2011tm}.
 While in \cite{esposito}, \cite{Moss:2003bk}  boundary conditions on the fields are imposed, in \cite{vanNieuwenhuizen:2005kg} it is pointed out that the supergravity action should be invariant under local supersymmetry \emph{without imposing Dirichlet boundary conditions on the fields}, in contrast to the Gibbons--Hawking prescription \cite{Gibbons:1976ue}. The explicit construction of an $AdS$ supergravity theory with a boundary and with no boundary conditions on the fields was achieved in reference \cite{vanNieuwenhuizen:2005kg} using superconformal tensor calculus, in the particular case of $N=1, D=3$ (off-shell) supergravity. Within that  approach,  it was shown in particular that $N=1$, $D=3$ pure supergravity, including its appropriate boundary term, actually reproduces not only the Gibbons-Hawking-York boundary term, but also the counterterm which regularizes the total action, in the language of holographic renormalization.
 An interesting geometric approach to the problem was considered, for the rigid superspace, in \cite{Howe:2011tm}, by applying the so-called ectoplasm formalism to manifolds with boundary. However, that approach seems to be related to the existence of an off-shell formulation of  supergravity which, as it is well known, is available only in few cases. Notwithstanding this, some of the statements in \cite{Howe:2011tm} appear to be consistent with the results we are going to present here.

Let us remark that the above results, together with the ones  of \cite{Aros:1999id}, all point to the conclusion that, to restore all the invariances of a gravity or supergravity lagrangian with cosmological constant in the presence of a non trivial boundary, it is necessary to add topological contributions,  also providing the counterterms needed to regularize the action and the conserved charges.
 \par
 In the present paper we work out the construction of $N=1$ and $N=2$, $D=4$ simple supergravities with negative cosmological constant and a non trivial boundary, thus generalizing to four dimensional  extended supergravity the results of \cite{Aros:1999id} and \cite{vanNieuwenhuizen:2005kg}.
To deal with this problem we take an approach  different from that of reference \cite{vanNieuwenhuizen:2005kg}, namely we introduce in a geometric way appropriate boundary terms to the lagrangian in such a way that the action, including the boundary contributions,  be invariant under supersymmetry transformations. In a sense our approach extends to superspace the geometric approach  of \cite{Aros:1999id}.
 As we are going to show, the geometric (or rheonomic) approach to supergravity, where the supersymmetry transformations are generated by Lie derivatives in the fermionic directions of superspace, seems particularly well suited for the completion of such a task. In particular, it does not require an off-shell formulation of bulk supergravity. \footnote{Indeed, we shall use for both $N=1$ and $N=2$ supergravity the  on-shell formulation (that is transformations close only on the equations of motion). However, our results can be easily extended to an off-shell formulation, when available. In the $N=1$ case this is easily performed using the auxiliary fields of the new minimal model. }

For pure gravity with cosmological constant the boundary term can be written as a purely topological addition to the space-time action, and it regularizes the boundary action without imposing Dirichlet boundary conditions on the metric.
As we will show, in the supergravity case the extra boundary terms that we introduce to recover full supersymmetry in the bulk and boundary of space-time, extend in a supersymmetric way the Euler density used in  \cite{Aros:1999id},  without imposing Dirichlet boundary conditions on the fields. On the contrary, we will find that  the boundary values of the field-strengths (more precisely of the supercurvatures) in superspace are  instead  dynamically fixed by the field equations of the full (bulk and boundary) lagrangian. As the Gauss-Bonnet term in pure gravity allows to recover invariance of the theory under all the bosonic symmetries, lost in the presence of a boundary of space-time, and further regularizes the action, it is tempting to argue that the same mechanism, in particular the generation of counterterms which regularize the action, should also be at work in the four-dimensional supersymmetric case.
  If it were the case, it would be very interesting to go further and evaluate the boundary contributions needed to restore supersymmetry in matter coupled and/or higher $N$-extended supergravity theories in 4 and higher dimensions.

\subsection{Our approach}

 Let us now clarify our geometrical approach for the description  of $N$-extended pure supergravity in four dimensions in the presence of a cosmological constant. Let  $V^a$ ($a=0,1,2,3$) and $\psi_A^\alpha$ ($A=1,\cdots N$, $\alpha=1,\cdots 4$) be the bosonic and fermionic vielbein 1-forms in superspace, respectively. The index $A$ is the $U(N)$ R-symmetry index while $\alpha$ is a $4D$ spinor index, that we will omit in the following.

The fundamental request of any supergravity theory is the invariance of the lagrangian $\mathcal L$ under supersymmetry transformations.
In the geometric approach, the theory is given in terms of superfields 1-forms $\mu^{\mathcal{A}}$ defined on superspace $\mathcal{M}_{4|4N}$.   The lagrangian,  as a functional of the $\mu^{\mathcal{A}}$, is   a bosonic 4-form in superspace and  the action is obtained by integrating $\mathcal L$ on a generic bosonic hypersurface $\mathcal M_4 (x,\theta)\subset \mathcal{M}_{4|4N} $ immersed in superspace.
\footnote{Note that within this geometric approach all the fields are superfields, but we never need to use an expansion in the fermionic $\theta$ coordinates. The space-time lagrangian is recovered at the end of the calculation by restricting all the fields and p-forms to their $\theta=0, d\theta=0$ content.}
In this setting, supersymmetry transformations in space-time are interpreted as diffeomorphisms in the fermionic directions of superspace leading from a given $\cM_4(x,\theta)$ to a nearby one $\cM_4(x,\theta +\delta\theta)$. They are generated by Lie derivatives with fermionic parameter $\epsilon^\alpha_A$. (See Appendix for more details).

It follows that supersymmetry invariance of the lagrangian is easily accounted for
 by asking that the Lie derivative $\ell_\epsilon$ of the lagrangian vanishes for infinitesimal diffeomorphisms in the fermionic directions.

 More precisely, let us denote by $\iota$ the contraction operator, and
 by $\epsilon_A(x,\theta)$ the fermionic parameter along the tangent vector $D^A$ dual to the gravitino $\psi_A$ ($\bar\psi_A^\alpha(D^B_\beta)=\delta_\beta^\alpha \delta_A^B$) and moreover $\iota_\epsilon(\psi_A)=\epsilon_A$, $\iota_\epsilon(V^a)=0$). The condition for the lagrangian to be invariant under local supersymmetry is:
\begin{equation}
\delta_\epsilon \mathcal{L}=\ell_\epsilon \mathcal L= \iota_\epsilon d\mathcal L + d(\iota_\epsilon \mathcal L)=0 \,.\label{lie}
\end{equation}
Let us note that the first contribution, which would be identically zero in space-time, is not trivial here, since $d\mathcal L$ is a 5-form \emph{in superspace}. The second contribution is a boundary term, that does not affect the bulk result. A necessary condition for a supergravity lagrangian is then:
\begin{equation}
\label{bulklie}\iota_\epsilon d\mathcal L=0\,,
\end{equation}
corresponding to require supersymmetry invariance in the bulk. We will assume in the following that the condition (\ref{bulklie})  always holds. Under this (necessary) condition, the supersymmetry transformation of the action reduces to:
\begin{equation}\label{varact}
\delta_\epsilon \mathcal{S}=\int_{\mathcal{M}_4} d(\iota_\epsilon \mathcal L)=\int_{\partial\mathcal{M}_4} \iota_\epsilon \mathcal L\,.
\end{equation}
When considering supergravity on Minkowski background, or more generally on space-times without a boundary, the fields are asymptotically vanishing so that $\left.\iota_\epsilon \mathcal L\right|_{\partial\mathcal{M}_4}=0$, and then $\delta_\epsilon \mathcal{S}=0$. In this case, eq. (\ref{bulklie}) is also a sufficient condition for the supersymmetry invariance of the lagrangian.

On the other hand, when the background space-time has a non trivial boundary, the condition
\begin{equation}\label{bound}
\left.\iota_\epsilon \mathcal L\right|_{\partial\mathcal{M}_4}= 0
 \end{equation}
 (modulo an exact differential)
 becomes non trivial, and it is necessary to check it explicitly to get supersymmetry invariance of the action.

For the cases we considered, we find that the bulk lagrangian $\cL_{bulk}$ is not supersymmetric when a boundary is present. In this case, we show that supersymmetry invariance is recovered by adding topological contributions $\cL_{bdy}$ to the bulk lagrangian.
 Even if they do not affect the bulk, they reestablish the supersymmetry invariance of the total lagrangian besides modifying the boundary dynamics.

Let us observe that  the total lagrangian  $\cL_{full}= \cL_{bulk}+\cL_{bdy}$ can be rewritten in a suggestive way  as a sum of quadratic terms in $OSp(N|4)$-covariant super-fieldstrengths. In particular, for the $N=1$ case our result reproduces the MacDowell--Mansouri action \cite{MacDowell:1977jt}. We extend this result to $N=2$ supergravity and we guess that the same structure should appear also for higher $N$ theories. The generalization of our results to matter coupled and/or  $N\geq 4$ theories, where also scalar fields are present, is in preparation.

\section{Pure $N=1$  supergravity in 4 dimensions}
In the $N=1$ theory the fermionic directions in superspace are spanned by one gravitino 1-form, $\psi$, which is a Majorana spinor.

The Lorentz-covariant field-strengths in superspace are:
\begin{eqnarray}\label{curv1}
\left\{\begin{array}{lll}
\mathcal R^{ab}&=& d\omega^{ab} - \omega^a_{\ c}\wedge \omega^{cb}\\
\rho&=& \mathcal D \psi \equiv d\psi -\frac 14 \omega^{ab}\gamma_{ab}\wedge\psi\\
R^a &=& \mathcal D V^a - \frac \ii 2 \bar\psi\gamma^a \wedge \psi\equiv (dV^a -\omega^a_{\ b}\wedge V^b ) - \frac \ii 2 \bar\psi\gamma^a \wedge \psi\end{array}\right.\,,
\end{eqnarray}
and they satisfy (on-shell) the Bianchi identities:
\begin{eqnarray}\label{bianchi1}
\left\{\begin{array}{lll}
\mathcal D \mathcal R^{ab}&=& 0\\
\mathcal D \rho&=& -\frac 14 \mathcal R^{ab}\gamma_{ab}\wedge\psi\\
\mathcal D R^a &=&  -\mathcal R^a_{\ b}\wedge V^b +   \ii   \bar\psi\gamma^a \wedge \rho \end{array}\right.\,.
\end{eqnarray}

Let us consider the following lagrangian in superspace, whose equations of motion admit an AdS$_4$ vacuum solution with cosmological constant $\Lambda = - 12 e^2$. The factor $e$ is related to the radius $\ell$ of the asymptotic AdS$_4$ geometry by: $e= \frac 1{2\ell}$.
\begin{eqnarray}
\label{lagrmink}
\mathcal L_{bulk}&=& -\frac 14\mathcal R^{ab}\wedge V^c \wedge V^d \epsilon_{abcd}- \bar \psi \gamma_5 \gamma_a \wedge \rho \wedge V^a+\nonumber\\
&&-\ii e \bar \psi \gamma_5 \gamma_{ab} \wedge \psi \wedge V^a\wedge V^b -\frac 12 e^2 V^a\wedge V^b\wedge V^c \wedge V^d \epsilon_{abcd}\,.
\end{eqnarray}
Note that it is written as a first-order lagrangian, and the field equation for the spin-connection $\omega^{ab}$ implies (up to boundary terms, which will be considered later) the vanishing, on-shell, of the supertorsion $R^a$ defined in eq. (\ref{curv1}).

The lagrangian (\ref{lagrmink})  is   on-shell invariant (in the bulk) under supersymmetry (according to \eq{bulklie}), that is $\iota_\epsilon(d\mathcal{L}_{bulk})=0$.

Even if not strictly necessary for the present discussion, in the following lines we are going to summarize in a few words, for the simple $N=1$ theory, the main issues of the geometric approach, with the aim to make contact with the formulation of supergravity in space-time.
In the geometric formalism we are using here, the condition that the theory (and in particular the lagrangian) is supersymmetry invariant is equivalent to the requirement that appropriate superspace constraints hold on-shell (see Appendix), and in particular:
\begin{equation}
\iota_\epsilon(\rho)= \ii e \gamma_a \epsilon V^a\,;\quad R^a=0
\end{equation}
 that is the gravitino 2-form should admit  on-shell the following parametrization on a basis of superspace
\footnote{Let us clarify that, here and in the following, the component of a field-strength along the bosonic vielbein is not its space-time component, but is instead what is called the supercovariant field-strength. Indeed, considering e.g. $\rho$, from (\ref{curv1}) and (\ref{paramrho}),
projecting along $dx^\mu \wedge dx^\nu$
we have that the space-time component $\rho_{\mu\nu}$ of $\rho$ is:
$$\rho_{\mu\nu}\equiv \mathcal D_{[\mu}\psi_{\nu]}= \rho_{ab}V^a_\mu V^b_\nu + \ii e \gamma_{[\nu} \psi_{\mu]}\,,$$
where $\rho_{ab}V^a_\mu V^b_\nu$ defines the supercovariant field strength of the gravitino.
}:
\begin{equation}\label{paramrho}
\rho= \rho_{ab}V^a \wedge V^b + \ii e \gamma_a \psi V^a\,.
\end{equation}
The Bianchi identities in superspace \eq{bianchi1} are solved by parametrizing (on-shell) the full set of field-strengths on a basis of superspace in the following way:
\begin{eqnarray}\label{param1}
\left\{\begin{array}{lll}
\mathcal R^{ab}&=& \mathcal R^{ab}{}_{cd}V^c\wedge V^d  + \bar\Theta^{ab}{}_c \psi V^c -e \bar \psi \gamma^{ab}\psi\\
\rho&=& \rho_{ab}V^a \wedge V^b + \ii e \gamma_a \psi V^a\\
R^a &=& 0\end{array}\right.\,,
\end{eqnarray}
where the spinorial superfield $\Theta^{ab}{}_c$ must be related to $\rho_{ab}$ by:
$\Theta_{ab|c}= \ii (2\gamma_{[a}\rho_{b]c}-\gamma_c \rho_{ab})
$.
Note that this parametrization can be equivalently read as the on-shell prescription for the contractions of the field strengths:
\begin{eqnarray}
\left\{\begin{array}{lll}
\iota_\epsilon(\mathcal R^{ab})&=&  \bar\Theta^{ab}{}_c \epsilon V^c -2 e \bar \epsilon \gamma^{ab}\psi\\
\iota_\epsilon(\rho)&=&  \ii e \gamma_a \epsilon V^a\\
\iota_\epsilon(R^a) &=& 0\end{array}\right.\,,
\end{eqnarray}
Let us observe that these constraints provide the supersymmetry transformation laws of the fields on space-time, under which the space-time lagrangian is invariant up to boundary terms (see equations (\ref{Lie1}), (\ref{lie2})).

\subsection{Including boundary terms}
As discussed in the introduction, for this theory the boundary invariance of the lagrangian under supersymmetry is not trivially satisfied and the condition (\ref{bound}) has to be checked explicitly. In fact we find that, if the fields do not vanish at the boundary
$$\left.\iota_\epsilon \mathcal L\right|_{\partial\mathcal{M}_4}\neq 0\,.$$
To restore supersymmetry invariance, it is possible to modify the lagrangian by adding boundary (topological) terms, which do not alter the bulk lagrangian only affecting the boundary lagrangian, so that (\ref{bulklie}) is still satisfied.

The only possible topological 4-forms terms compatible with the symmetries of the theory (parity, Lorentz-invariance) are:
\begin{eqnarray}
  &&d\left(\mathcal \omega^{ab}\wedge  \mathcal{R}^{cd} - \omega^{a}{}_\ell \wedge \omega^{\ell b} \wedge  \omega^{cd}\right)\epsilon_{abcd}= \mathcal{R}^{ab}\wedge \mathcal{R}^{cd}\epsilon_{abcd}\label{euler1}
  \\&& d(\bar\psi\wedge \gamma^5 \rho)= {\bar\rho}\gamma_5
 \rho + \frac 14 \mathcal{R}^{ab} \bar\psi\gamma_5 \gamma_{ab}\psi
  \label{euler2}
\end{eqnarray}
corresponding to the boundary lagrangian:
 \begin{eqnarray}\label{lagb}
 \mathcal L_{bdy}&=&\alpha \mathcal R^{ab}\wedge \mathcal R^{cd}\epsilon_{abcd}+ \beta  \left({\bar\rho}\gamma_5
 \rho + \frac 14 \mathcal{R}^{ab} \bar\psi\gamma_5 \gamma_{ab}\psi \right)
\end{eqnarray}
where $\alpha$ and $\beta$ must be proportional to $e^{-2}$ and $e^{-1}$ respectively in order to respect the common scaling behaviour of all the terms of the lagrangian.
Let us then consider the following lagrangian
 \begin{eqnarray}\label{lagfull}
 \mathcal L_{full}&=&\mathcal L_{AdS}+\mathcal L_{bdy}\nonumber\\
 &=& -\frac 14\mathcal R^{ab}\wedge V^c \wedge V^d \epsilon_{abcd}- \bar \psi \gamma_5 \gamma_a \wedge \rho \wedge V^a+\nonumber\\
&&-\ii e \bar \psi \gamma_5 \gamma_{ab} \wedge \psi \wedge V^a\wedge V^b -\frac 12 e^2 V^a\wedge V^b\wedge V^c \wedge V^d \epsilon_{abcd}+\nonumber\\
&&\alpha \mathcal R^{ab}\wedge \mathcal R^{cd}\epsilon_{abcd}+ \beta  \left({\bar\rho}\gamma_5
 \rho + \frac 14 \mathcal{R}^{ab} \bar\psi\gamma_5 \gamma_{ab}\psi \right)
\end{eqnarray}
Recalling the general discussion in the introduction, and in particular (\ref{lie}), let us now study the conditions under which (\ref{lagfull}) is invariant under supersymmetry. Since the boundary terms (\ref{euler1}) and (\ref{euler2}) are total differentials,  the condition for supersymmetry in the bulk, $\iota_\epsilon d\mathcal L_{full}=0$, is trivially satisfied.
To prove the supersymmetry invariance of $\mathcal L_{full}$, it is still to be proved (from (\ref{lie})) that, for a suitable choice of $\alpha$ and $\beta$, $\iota_\epsilon(\mathcal L_{full})$ vanishes on the boundary (up to a total derivative).
 We have:
 \begin{eqnarray}\label{iotal}
 \iota_\epsilon(\mathcal L_{full})&=& -\frac 14\iota_\epsilon(\mathcal R^{ab})\wedge V^c \wedge V^d \epsilon_{abcd}- \bar \epsilon \gamma_5 \gamma_a   \rho \wedge V^a  + \bar \psi \gamma_5 \gamma_a \wedge \iota_\epsilon(  \rho) \wedge V^a + \nonumber\\
 &&-2\ii e \bar \epsilon \gamma_5 \gamma_{ab} \psi V^a V^b+\nonumber\\
 &&+  2 \iota_\epsilon(\mathcal R^{ab})\left(\alpha\mathcal R^{cd}- \frac \ii 8 \beta \bar\psi\gamma^{cd}\psi\right)\epsilon_{abcd}+ 2\beta  \iota_\epsilon({\bar\rho})\gamma_5\rho -\frac \ii 4 \beta \mathcal R^{ab}\wedge  \bar\epsilon\gamma^{cd}\psi \epsilon_{abcd}
  \nonumber\\
 \end{eqnarray}
This is not zero, in general, but its projection on the boundary $\partial \mathcal M$ should be zero.  Indeed, in the presence of a boundary, the field equations in superspace of the lagrangian (\ref{lagfull}) acquire non trivial boundary contributions, which result in the following constraints holding on the boundary:\footnote{Note that besides the contributions to the equations of motion coming from $\mathcal{L}_{bdy}$ we have also extra contributions from $\mathcal{L}_{bulk}$ (neglected in the absence of a boundary), from the total differentials originating from partial integration.}
\begin{eqnarray}\label{bdyconstraints}
\left\{\begin{array}{lll}
\frac{\delta \mathcal L_{full}}{\delta \omega^{ab}} =0 &\Rightarrow & \mathcal R^{ab}|_{\partial \mathcal M}= \frac 1{8\alpha} V^a V^b + \ii \frac \beta{16 \alpha} \bar\psi\gamma^{ab}\psi\\
\frac{\delta \mathcal L_{full}}{\delta \bar \psi} =0 &\Rightarrow & \rho|_{\partial \mathcal M} = \frac{1}{2\beta} \gamma_a \psi V^a
\end{array}\right.\,,
\end{eqnarray}
which show that both supercurvatures on the boundary are not  dynamical, but fixed to constant values in the anholonomic basis of the bosonic and fermionic vielbein.
Upon use of (\ref{bdyconstraints}) we then find that
\begin{eqnarray}\label{iotal'}
 \iota_\epsilon(\mathcal L_{full})|_{\partial \mathcal M}&=& 0\,,
 \end{eqnarray}
 if the following relation between $\alpha$ and $\beta$ holds:

 \begin{equation}\label{rela}
   \frac{\beta}{16\alpha} +\frac{1}{2\beta}=2\ii\, e
\end{equation}
Solving for $\beta$ we find

$$
\beta=16\ii\,e\alpha\left( 1 + k\right),\quad k^2=1 + \frac{1}{32\,e^2\alpha}\,;\quad\quad  (\beta\neq 0 \Rightarrow k\neq -1).
$$
It is interesting to
 observe that setting $k=0$, what implies
 $$
 \alpha= -\frac{1}{32\,e^2}
 $$
  the full lagrangian can be written in terms of the $OSp(1|4)$ covariant super curvatures: 
\begin{eqnarray}\label{curv2}
\left\{\begin{array}{lll}
 \hat R^{ab}&=& \mathcal R^{ab} + 4 e^2 V^a \wedge V^b + e \bar\psi \gamma^{ab}\psi\\
\hat \rho&=& \rho -\ii e \gamma_a \psi \wedge V^a\\
R^a &=& \mathcal D V^a - \frac \ii 2 \bar\psi\gamma^a \wedge \psi\end{array}\right.\,,
\end{eqnarray}
satisfying (on-shell) the Bianchi identities (totally equivalent to \eq{bianchi1}):
\begin{eqnarray}\label{bianchi2}
\left\{\begin{array}{lll}
\mathcal D \hat R^{ab}&=& 8 e^2 R^{[a}\wedge V^{b]}-2 e \bar \psi \gamma^{ab}\hat \rho\\
\mathcal D \hat \rho&=& -\frac 14 \hat R^{ab}\gamma_{ab}\wedge\psi + \ii e \gamma _a \psi\wedge R^a -\ii e \gamma_a \hat\rho \wedge V^a\\
\mathcal D R^a &=&  - \hat R^a_{\ b}\wedge V^b +   \ii   \bar\psi\gamma^a \wedge \hat \rho \end{array}\right.\,.
\end{eqnarray}
In terms of (\ref{curv2}) and setting $k=0$, the lagrangian (\ref{lagfull}) turns out to be:
 \begin{eqnarray}\label{lagfullads}
 \mathcal L_{full}&=&-\frac 1{32e^2}  \hat R^{ab}\wedge \hat  R^{cd} \epsilon_{abcd} - \frac \ii{2e }  \hat{\bar \rho}\gamma_5
 \hat \rho\,.
\end{eqnarray}
This is in fact nothing but the Mac Dowell--Mansouri action \cite{MacDowell:1977jt}. We note that in terms of the supercurvatures (\ref{curv2}), the boundary conditions on the field-strengths (\ref{bdyconstraints}) take the simple form $ \hat{ \mathcal{R}}^{ab}|_{\partial \mathcal M}=0$ and $\hat \rho|_{\partial \mathcal M} =0$, that is we find that the $Osp(1|4)$ supercurvatures vanish at the boundary.\footnote{The limit case of a vanishing cosmological constant ($e=0$) in the presence of a non trivial boundary of space-time is interesting and will be considered elsewhere.}
We remark that the value of $\alpha= -\frac 1{32 e^2}$ we find for $k=0$ is precisely the one given  in \cite{Aros:1999id} where the invariance of the gravity lagrangian under space-time diffeomorphisms was required. Our result gives the supersymmetric extension of their arguments. 

On the other hand, taking $k\neq 0$ amounts to adding to the the Mac Dowell--Mansouri  lagrangian the boundary terms
\begin{equation}\label{kappa}
\frac{k^2}{32 e^2(k^2-1)}d\left(\mathcal \omega^{ab}\wedge  \mathcal{R}^{cd} - \omega^{a}{}_\ell \wedge \omega^{\ell b} \wedge  \omega^{cd}\right)\epsilon_{abcd}+ 16\ii\,e\alpha k\, d\,(\overline\psi\gamma^5\wedge \rho).
\end{equation}
These terms break the off-shell $OSp(1|4)$ structure of the theory. 
However, as discussed above the first term is incompatible with the invariance of the lagrangian under diffeomorphisms in the bosonic directions of superspace (as discussed in \cite{Aros:1999id}). As far as the second term is considered, 
using the value of $\rho$ at the boundary, (\ref{bdyconstraints}), we see that this term vanishes identically on-shell:
\begin{equation}\label{van}
 (\overline\psi\wedge\gamma^5 \rho)|_{\partial \mathcal M}=   \overline\psi\wedge\gamma^5\gamma_a\psi\wedge V^a\equiv 0.
\end{equation}
In view of the fact that the closure of the $OSp(1|4)$ algebra only holds on-shell for superymmetric theories (without auxiliary fields), this extra contribution does not play a significant role as far as supersymmetry is concerned.\footnote{As we shall see in the following, in the $N=2$ case all the coefficients of the boundary terms are fixed and no possible $OSp(2|4)$ breaking term can appear at the boundary.}\\
 In the next section, following the same arguments, we are going to extend the above construction  to $N=2$ pure supergravity.

\section{Pure $N=2$  supergravity in 4 dimensions}
The bulk contribution to $N=2$ supergravity on a AdS background can be found from the general matter coupled $N=2$ supergravity of \cite{Andrianopoli:1996cm}, by setting to zero the matter supermultiplets still keeping a Fayet--Iliopoulos (FI) term in the hypermultiplet sector. To do the calculation, we chose the FI term in a specific $SU(2)$ direction, $P = P^{(x=2)}_{\Lambda=0}$ ($x$ being a three-dimensional vector index) thus breaking the manifest $SU(2)$ invariance of the theory to $SO(2)$.
To precise the notation, we decompose the $N=2$ gravitinos in chiral components, the position of the $SU(2)$ index $A=1,2$ also denoting the chirality: $$\psi_A \equiv \frac{1+\gamma_5}2 \psi_A \,,\quad \psi^A \equiv \frac{1-\gamma_5}2 \psi^A $$
so that $\psi_A$ are left-handed gravitinos while $\psi^A$ are right-handed.

The lagrangian is written in a first order approach  for the spin connection $\omega^{ab}$ and also for the gauge field (see Appendix).  In the absence of matter fields, the kinetic matrix $\mathcal{N}_{\Lambda\Sigma}$ of the gauge field strength $\mathcal F= d A$ reduces to $\mathcal{N}_{00}=\theta -\ii$, where $\theta $ is the (constant) $\theta$-angle and we have set the gauge coupling constant $g=1$. The resulting (bulk) lagrangian, written as a 4-form in $N=2$ superspace, is:
\begin{eqnarray}\label{n2bulk}
\mathcal{L}_{bulk}&=& -\frac 14 \mathcal R^{ab}\wedge V^c \wedge V^d \epsilon_{abcd}+ \left(\bar \psi^A   \gamma_a \wedge \rho_A -\bar \psi_A   \gamma_a \wedge \rho^A\right)\wedge V^a+\nonumber\\
&&+\left(\theta \tilde F_{ab}+ \frac 12 \epsilon_{abcd}\tilde F^{cd}\right)V^a\wedge  V^b
\wedge F+ \nonumber\\
&&-\frac 1{24} \left( \tilde F_{\ell m}\tilde F^{\ell m} -\frac \theta 2 \epsilon_{pqrs}\tilde F^{pq}\tilde F^{rs}\right)V^a\wedge V^b \wedge V^c \wedge V^d \epsilon_{abcd}\nonumber\\
&&-L\left[ F -\frac L2\left(\bar\psi^A\psi^B\epsilon_{AB}+\bar\psi_A\psi_B\epsilon^{AB}\right)\right]\left[(\theta -\ii)\bar\psi^C\psi^D\epsilon_{CD}+(\theta +\ii)\bar\psi_C\psi_D\epsilon^{CD}\right]\nonumber\\
&&+\ii \left(S_{AB} \bar \psi^A  \gamma_{ab} \wedge \psi^B - \bar S^{AB}\bar\psi_A \gamma_{ab} \psi_B\right)\wedge V^a\wedge V^b
\nonumber\\&&- \frac 18 L^2 P^2  V^a\wedge V^b\wedge V^c \wedge V^d \epsilon_{abcd}\,,
\end{eqnarray}
where $L\equiv L^0$ is a constant which is the remnant of the special geometry section $L^\Lambda$ in the absence of scalars (with our notations
$L=\frac 1{\sqrt{2}}$), and the gravitino mass matrix is:
\begin{eqnarray}\label{shift}
S_{AB}= \frac \ii 2 (\sigma^x)_A{}^C \epsilon_{BC} P^x_\Lambda L^\Lambda = \frac 12 PL \delta_{AB}\,,\quad \bar S^{AB}= \frac \ii 2 (\sigma^x)^A{}_B \epsilon^{BC} P^x_\Lambda L^\Lambda = \frac 12 PL \delta^{AB}\,.
 \end{eqnarray}
 Correspondingly, the cosmological constant is $\Lambda = - 3 L^2 P^2$, and the radius $\ell$ of the asymptotic AdS$_4$ geometry is $\ell= \frac 1{LP}$.

Note that the lagrangian \eq{n2bulk} contains a $\theta$-term which is in fact a boundary contribution.
The super curvatures appearing in the lagrangian are defined as follows:
\begin{eqnarray}
\left\{\begin{array}{lll}
\mathcal{R}^{ab}&=& d\omega^{ab} -\omega^{ac}\wedge\omega_c{}^b\\
\rho_A&=& \mathcal{D}\psi_A + \frac \ii 2 A P_A{}^B \psi_B\\
\rho^A&=& \mathcal{D}\psi^A -  \frac \ii 2 A P^A{}_B \psi^B\\
F&=&  \mathcal{F} + L\left(\bar\psi^A\psi^B\epsilon_{AB}+\bar\psi_A\psi_B\epsilon^{AB}\right)\\
R^a&=& dV^a -\ii \bar\psi_A \gamma^a \psi^A\end{array}\right.
\end{eqnarray}
where $\mathcal{F}\equiv dA$ and we have defined $P_A{}^B\equiv (\sigma^2)_A{}^B P= P^A{}_B $.
They satisfy the $N=2$ Bianchi identities:
\begin{eqnarray}\label{N2BI}
\left\{\begin{array}{lll}\mathcal{D}
\mathcal{R}^{ab}&=& 0\\
\nabla \rho_A&\equiv& \mathcal{D}\rho_A + \frac \ii 2 A P_A{}^B \rho_B= -\frac 14\mathcal{R}^{ab}\gamma_{ab}\psi_A + \frac \ii 2 \mathcal{F} P_A{}^B \psi_B \\
\nabla \rho^A&\equiv& \mathcal{D}\rho^A - \frac \ii 2 A P^A{}_B \rho^B= -\frac 14\mathcal{R}^{ab}\gamma_{ab}\psi^A - \frac \ii 2 \mathcal{F} P^A{}_B \psi^B \\
d F&=&  2 L\left(\bar\psi^A\rho^B\epsilon_{AB}+\bar\psi_A\rho_B\epsilon^{AB}\right)\\
\mathcal{D}R^a&=& -\mathcal{R}^{ab}V_b +\ii \left(\bar\psi_A \gamma^a \rho^A+\bar\psi^A \gamma^a \rho_A\right)\end{array}\right.
\end{eqnarray}
where $\mathcal{D}$ denotes the Lorentz-covariant differential.

According to eq. \eq{lie},
the lagrangian (\ref{n2bulk}) is invariant under supersymmetry up to boundary terms, since
\begin{equation}
\iota_\epsilon d\mathcal{L}_{bulk}=0
\end{equation}
where $\epsilon \equiv \bar \epsilon_A D^A + \bar\epsilon^A D_A$ is the parameter corresponding to an infinitesimal supersymmetry transformation, namely a Lie derivative under an infinitesimal diffeomorphism in the $\theta^\alpha$ fermionic directions of superspace. The tangent vectors ${D^A,D_A}$ are dual to the gravitino 1-forms $\bar\psi^A(D_B)=\bar\psi_A(D^B)=\delta^A_B$, and moreover we have: $\iota_\epsilon \psi_A= \epsilon_A$, $\iota_\epsilon \psi^A= \epsilon^A$, $\iota_\epsilon V^a= 0$.
In the presence of a boundary where the fields do not vanish (as is the case for supergravity in AdS background),
  the action looses its  supersymmetry invariance, since, from (\ref{varact}), we still have a non vanishing contribution:
  \begin{equation}
  \delta_\epsilon \mathcal{S}=\int_{\mathcal{M}_4} d (\iota_\epsilon \mathcal{L}_{bulk}) =  \int_{\partial\mathcal{M}_4}  \iota_\epsilon \mathcal{L}_{bulk}\neq 0\,.
  \end{equation}
  The lack of boundary supersymmetry invariance is also manifest from the fact that in this case the boundary contributions to the field equations are not satisfied anymore, since in particular:
  \begin{equation}
  \label{boundem}
  \frac{\delta \mathcal{L}_{bulk}}{\delta \bar \psi}|_{\partial \mathcal{M}}\neq 0\,,\quad \frac{\delta \mathcal{L}_{bulk}}{\delta A}|_{\partial \mathcal{M}}\neq 0\,,\quad \frac{\delta \mathcal{L}_{bulk}}{\delta \omega}|_{\partial \mathcal{M}}\neq 0\,.
\end{equation}
To restore full supersymmetry in the bulk and boundary, it is necessary to add topological contributions.
The possible boundary contributions  that could be added to the lagrangian are:
 \begin{eqnarray}
 \mathcal{L}_{bdy}&=& d\left\{\alpha (\omega^{ab}\wedge  \mathcal{R}^{cd} - \omega^{a}{}_\ell \wedge \omega^{\ell b} \wedge  \omega^{cd})\epsilon_{abcd}+\right.
  \nonumber\\
 &&
\left.+\beta S_{AB} \bar\psi^A \rho^B+\bar \beta {\bar S}^{AB} {\bar\psi}_A \rho_B +
\gamma A\mathcal{F} \right\}
 \end{eqnarray}
 which can be written, using \eq{N2BI}:
  \begin{eqnarray}
 \mathcal{L}_{bdy}&=& \alpha \mathcal{R}^{ab}\wedge  \mathcal{R}^{cd} \epsilon_{abcd}+\beta S_{AB} \bar\rho^A \rho^B+\bar \beta {\bar S}^{AB} {\bar\rho}_A \rho_B +
  \gamma \mathcal{F}\wedge\mathcal{F}
+\nonumber\\
&&+ \frac 14 \mathcal{R}^{ab} \left(\beta S_{AB} \bar\psi^A \gamma_{ab}\psi^B+\bar \beta {\bar S}^{AB} {\bar\psi}_A\gamma_{ab}\psi_B\right)+\nonumber\\
&&+\frac \ii 2 \mathcal{F}
\left(\beta S_{AB}P^B{}_C \bar\psi^A \psi^C-\bar \beta {\bar S}^{AB} P_B{}^C{\bar\psi}_A\psi_C\right)
 \end{eqnarray}
 Let us then consider
 \begin{eqnarray}
 \mathcal{L}_{full}=\mathcal{L}_{bulk}+\mathcal{L}_{bdy}\,.
 \end{eqnarray}
 The addition of the boundary terms modifies the field equations on the boundary. The request $\frac{\delta \mathcal{L}_{bulk}}{\delta \mu^A}|_{\partial \mathcal{M}}=0$  (where $\mu^A=\{\omega^{ab},V^a, A,\psi_A,\psi^A\}$  generically denotes all the super fields of the theory) can now be satisfied, and it  implies the following constraints:
 \begin{eqnarray}
 \mathcal{R}^{ab}|_{\partial \mathcal{M}} &=& \frac 1{8\alpha}\left\{ V^a V^b - \frac \ii 2 \left(\beta  S_{AB} \bar\psi^A \gamma_{ab}\psi^B-\bar \beta {\bar S}^{AB} {\bar\psi}_A\gamma_{ab}\psi_B\right)\right\}
 \\
 {F}|_{\partial \mathcal{M}} &=& -\frac 1{2\gamma}\left\{\left(\theta \tilde F_{ab}+ \frac 12 \epsilon_{abcd}\tilde F^{cd}\right)V^a\wedge  V^b +\right.\nonumber\\
 &&\hskip -3mm\left. -L\left[(\theta  +2 \gamma -\ii +\beta \frac {P^2}{4})\epsilon_{AB}\bar\psi^A\psi^B+
 (\theta+2\gamma  +\ii +\bar\beta \frac {P^2}{4})\epsilon^{AB}\bar\psi_A\psi_B\right]\right\}\\
 \rho_A|_{\partial\mathcal{M}}&=& \frac 1{2\bar\beta} (\bar S^{-1})_{AB} \gamma_a \psi^B V^a\\
 \rho^A|_{\partial\mathcal{M}}&=& -\frac 1{2\beta} ( S^{-1})^{AB} \gamma_a \psi_B V^a \,.
 \end{eqnarray}
Let us now ask the full supersymmetry invariance of the total lagrangian, that is the condition \eq{lie}. The condition $\iota_\epsilon (d\mathcal{L}_{full})=0$ coincides with the request of bulk supersymmetry invariance, since $d\mathcal{L}_{bdy}\equiv 0$.
To have also boundary supersymmetry invariance, we must fix the parameters $\alpha,\beta,\bar\beta,\gamma$ to particular values, and we find:
\begin{equation}
\iota_\epsilon \left(\mathcal{L}_{full}\right)_{\partial \mathcal{M}}=0 \quad \Rightarrow \quad\alpha= -\frac 1{4P^2}\,,\quad \beta = \ii \frac{4}{P^2}=-\bar\beta\,,\quad \gamma =-\frac 12\theta
\end{equation}
With the above values the total lagrangian becomes:
\begin{eqnarray}\label{n2full}
\mathcal{L}_{full}&=& -\frac 14 \mathcal R^{ab}\wedge V^c \wedge V^d \epsilon_{abcd}+ \left(\bar \psi^A  \gamma_a \wedge \rho_A -\bar \psi_A \gamma_a \wedge \rho^A\right)\wedge V^a+\nonumber\\
&&+\frac 12 \epsilon_{abcd}\tilde F^{cd}V^a\wedge  V^b
\wedge F-
\frac 1{24}  \tilde F_{\ell m}\tilde F^{\ell m} V^a\wedge V^b \wedge V^c \wedge V^d \epsilon_{abcd}\nonumber\\
&&+\ii L\left[ F -\frac L2\left(\bar\psi^A\psi^B\epsilon_{AB}+\bar\psi_A\psi_B\epsilon^{AB}\right)\right]\left[\bar\psi^C\psi^D\epsilon_{CD}-\bar\psi_C\psi_D\epsilon^{CD}\right]\nonumber\\
&&+\ii \left(S_{AB} \bar \psi^A  \gamma_{ab} \wedge \psi^B - \bar S^{AB}\bar\psi_A \gamma_{ab} \psi_B\right)\wedge V^a\wedge V^b
\nonumber\\&&+ \frac 12 L^2 P^2  V^a\wedge V^b\wedge V^c \wedge V^d \epsilon_{abcd}+\nonumber\\
&&-\frac 1{4P^2} \left\{\mathcal{R}^{ab}\wedge  \mathcal{R}^{cd} \epsilon_{abcd}- 16\ii (S_{AB} \bar\rho^A \rho^B- {\bar S}^{AB} {\bar\rho}_A \rho_B) +\right.
\nonumber\\
&&-4\ii  \mathcal{R}^{ab} \left( S_{AB} \bar\psi^A \gamma_{ab}\psi^B- {\bar S}^{AB} {\bar\psi}_A\gamma_{ab}\psi_B\right)+\nonumber\\
&&\left.+8 \mathcal{F}
\left(S_{AB}P^B{}_C \bar\psi^A \psi^C+ {\bar S}^{AB} P_B{}^C{\bar\psi}_A\psi_C\right)\right\}\,.
\end{eqnarray}
We note in particular that \emph{the value of $\gamma=-\frac 12 \theta$ is such as to exactly cancel the topological $\theta$-term in the gauge sector.}

 The boundary values of the supercurvatures become:
  \begin{eqnarray}\label{curvbdy2}
\left\{\begin{array}{lll} \mathcal{R}^{ab}|_{\partial \mathcal{M}} &=& -\frac {P^2}2\left\{ V^a V^b + \frac2{P^2} \left(  S_{AB} \bar\psi^A \gamma_{ab}\psi^B+ {\bar S}^{AB} {\bar\psi}_A\gamma_{ab}\psi_B\right)\right\}
 \\
 {F}|_{\partial \mathcal{M}} &=&  0 \\
 \rho_A|_{\partial\mathcal{M}}&=& -\frac {P^2}{8}\ii  (\bar S^{-1})_{AB} \gamma_a \psi^B V^a\\
 \rho^A|_{\partial\mathcal{M}}&=& \frac {P^2}{8}\ii  ( S^{-1})^{AB} \gamma_a \psi_B V^a
\end{array}\right. \,.\end{eqnarray}

As it happens in the $N=1$ case, the total lagrangian \eq{n2full} simplifies dramatically when written  in terms of the  $OSp(2|4)$-curvatures defined below:
 \begin{eqnarray}\label{lagsuper}
\left\{\begin{array}{lll}
\hat{{R}}^{ab}&\equiv&{\mathcal{R}}^{ab} + \frac{P^2}2 V^a V^b + S_{AB} \bar\psi^A \gamma_{ab}\psi^B+ {\bar S}^{AB} {\bar\psi}_A\gamma_{ab}\psi_B\\
\hat\rho_A&\equiv&\rho_A  -   \ii S_{AB}\gamma_a \psi^B V^a\\
\hat\rho^A&\equiv& \rho^A -   \ii \bar S^{AB}\gamma_a \psi_B V^a\\
F&\equiv&  \mathcal{F} + L\left(\bar\psi^A\psi^B\epsilon_{AB}+\bar\psi_A\psi_B\epsilon^{AB}\right)
\end{array}\right.\end{eqnarray}
Indeed one can easily verify that $\cL_{full}$ can be written in the quite simpler  form:
\begin{eqnarray}\label{lagAdS}
\mathcal{L}_{full}&=& -\frac 1{4 P^2}\hat R^{ab}\hat R^{cd}\epsilon_{abcd}+\frac{4}{P^2}\ii \left(S_{AB}\hat{\bar\rho}^A\hat\rho^B -\bar S^{AB}\hat{\bar\rho}_A\hat\rho_B\right)+\nonumber\\
&&+\frac 12\left( \epsilon_{abcd}\tilde F^{cd}V^a\wedge  V^b\wedge F -\frac 1{12} \tilde F_{\ell m}\tilde F^{\ell m} V^a\wedge V^b \wedge V^c \wedge V^d \epsilon_{abcd}\right)\,.
\end{eqnarray}
Note that using (\ref{shift}) and recalling $L=1/\sqrt 2$, comparison between equations (\ref{curvbdy2}) and (\ref{lagsuper}) shows that all the $OSp(2|4)$-curvatures vanish at the boundary.\\

As explained shortly in the introduction and more extensively in Appendix A, the lagrangian (\ref{lagAdS}) is a bosonic 4-form embedded in superspace written in terms of supercurvatures and superfields. To obtain the lagrangian in ordinary space-time it is sufficient to set the fermionic coordinates $\theta^\alpha=0$ and their differential $d\theta^\alpha=0$ so that the hypersurface is identified with space-time. Moreover we can go from the first order to second order formalism for the gauge field-strength, that is we set $\tilde F_{ab}=F_{ab}$ (see Appendix A). In this way we obtain the space-time lagrangian:

\begin{eqnarray}\label{lag}
\mathcal{L}_{full}^{(space-time)}&=& -\frac 1{4 P^2}\hat R^{ab}\wedge\hat R^{cd}\epsilon_{abcd}+\frac{4}{P^2}\ii \left(S_{AB}\hat{\bar\rho}^A\wedge\hat\rho^B -\bar S^{AB}\hat{\bar\rho}_A\wedge\hat\rho_B\right)+\nonumber\\
&&+  \frac12 F\wedge {}^*F
\end{eqnarray}
where now all fields and curvatures are purely space-time 1-forms and 2-forms, respectively.

\section*{Acknowledgements}
We are grateful to R. Olea for having addressed our attention to the boundary problem in gravity  and supergravity and for illuminating discussions. We further thank M. Trigiante and A. Santambrogio for interesting discussions and criticism.
\appendix
\section{Derivation of the space-time lagrangian from
the geometric approach}
\label{appendiceA}
\setcounter{equation}{0}
\addtocounter{section}{0}

In this Appendix we give a short account of how to recover a supersymmetric lagrangian on space-time from a geometrical lagrangian in superspace according to the principle of rheonomy.  By a geometrical lagrangian we mean a lagrangian constructed using only $p$-forms, wedge products and $d$-differential.

In the geometric (rheonomic) approach the $p$-form fields $\mu^A$ are extended from space-time to superspace, $\mu^A(x)\rightarrow \mu^A(x,\theta)$. The lagrangian  is a bosonic
4-form in superspace integrated on a 4-dimensional (bosonic)
hypersurface ${\cal M}^{  4}$ locally embedded in superspace ${\cal M}^{ {4}\vert  {N}}$, to get the action:
\begin{equation}
{\cal S} = \int_{{\cal M}^{4} \subset {\cal M}^{4  \vert  n}} \, {\cal L}\,.
\label{campodicalcio}
\end{equation}
The field equations,  derived from the generalized variational principle
$\delta \cal S = {\rm{0}}$, are taken to hold in all superspace.  They are independent of the
particular hypersurface $\cal M^{\rm 4}$ on which we integrate, since any variation of the hypersurface can be reabsorbed in a superspace diffeomorphism. However, the notion of Hodge operator is not well defined in superspace.  The unavailability of the Hodge star-operator, whose presence would prevent the extension of the $p$-form equations to the whole superspace, forbids the possibility of writing the kinetic term of gauge fields in the lagrangian as $\mathcal F\wedge *\mathcal F$, but the kinetic term can be recovered in a simple way using a first order formalism for the gauge field. Actually one introduces a 0-form field $\tilde F_{ab}$ in such a way that its equations of motion perform the identification $\tilde F_{ab}= F_{ab}$, $F_{ab}$ being the bosonic-vielbein component of the field strength 2-form. To see explicitly how it works let us consider the following lagrangian
\begin{equation}\label{Ffirstorder}
\mathcal{L}=
\int\left(\frac 12 \epsilon_{abcd}\tilde F^{cd}\right)F\wedge V^a\wedge  V^b
+
 a \int\left( \tilde F_{\ell m}\tilde F^{\ell m} \right)V^a\wedge V^b \wedge V^c \wedge V^d \epsilon_{abcd}
\end{equation}
Varying the tensor 0-form $\tilde F_{cd}$ and choosing $a=-\frac{1}{24}
$ we find $\tilde F_{cd}= F_{cd}$. Inserting this result in \eq{Ffirstorder}
we obtain the desired result, namely
\begin{equation}\label{desiree}
  \mathcal{L} =\frac{1}{2} \int F\wedge {}^*F\,.
\end{equation}
Note that the first order formalism for the gauge field lagrangian is quite analogous to the first order formalism for the spin connection for the Einstein gravity term.

There are simple rules which can be used in order to write down
the most general lagrangian compatible with these requirement.
The implementation of these rules is described in detail in the
literature to which we refer the interested reader \cite{librone}.
Let us just summarize here the main argument which connects superspace diffeomorphisms to space-time supersymmetry.

The mapping $\mu^\mathcal{A}(x) \to \mu^\mathcal{A}(x,\theta)$ is defined by the requirement of
 \emph{rheonomy}, which amounts to the following:

The superspace equations of motion contain the superspace curvatures, and can be analyzed along the basis of 2-forms in superspace. Rheonomy
requires that expanding the curvatures 2-forms in superspace along the supervielbein 2-forms:
\begin{equation}\label{exp}
   R^\mathcal{A}= R^\mathcal{A}_{ab}V^a\wedge V^b+R^\mathcal{A}_{a \alpha}V^a\wedge \psi^\alpha+R^\mathcal{A}_{ \alpha\beta}\psi^\alpha\wedge \psi^\beta
 \end{equation}
 the "outer" components of $R^\mathcal{A}_{a \alpha}, R^\mathcal{A}_{ \alpha\beta}$  (namely the components in the fermionic directions, with at least 1 gravitino $\psi^\alpha $) be expressed as linear tensor combinations of the inner components $R^\mathcal{A}_{ab}$ (namely the components along the bosonic vielbeins $V^a$). Note that from a physical point of view the rheonomy requirement  avoids the introduction the spurious degrees of freedom which would appear in the theory if the outer components of the curvatures where independent fields.

  The rheonomy principle  is completely equivalent to the requirement of  space-time supersymmetry.
  Indeed let us recall that the Lie derivative formula $ \ell_\epsilon \mu^\mathcal{A}=\iota_\epsilon d \mu^\mathcal{A}+ d\iota_\epsilon \mu^\mathcal{A}$ can be written in the alternative form
\begin{equation}\label{Lie}
    \ell_\epsilon \mu^\mathcal{A}=(\nabla\epsilon)^\mathcal{A} + \iota _\epsilon R^\mathcal{A}
  \end{equation}
   where $\nabla$  denotes the covariant derivative of the parameter $\epsilon^\mathcal{A}$ corresponding to a gauge transformation. If rheonomy holds, the term $\iota_\epsilon R^\mathcal{A}$ will be given in terms of the components of the curvatures along the bosonic vielbein only. Therefore a transformation under diffeomorphisms in the $\theta^\alpha $ directions of superspace where only the fermionic parameter $\epsilon^\alpha$ is non-vanishing, takes the form
\begin{equation}\label{Lie1}
    \delta \mu^\mathcal{A}= \mu^\mathcal{A}(x,\theta+\delta \theta)- \mu^\mathcal{A}(x,\theta)=(\nabla \epsilon)^\mathcal{A}+\epsilon^\alpha C_{\alpha|\mathcal{A}}^{ab}R^\mathcal{A}_{ab}.
   \end{equation}
   where $C_{\alpha|\mathcal{A}}^{ab}$ is a constant tensorial quantity.
   This is a passive point of view for the Lie derivative, a flow from one hypersurface to another one translated by $\delta\theta$ \footnote{This flow from one hypersurface to another is responsible of the name ``\textit{rheonomy}'', which comes from the Greek  words $`\rho\epsilon \hat\iota\nu$ $\to$ flow and $\nu$\'o$\mu$os $\to$ law.}.  We can take however an active point of view, sticking to the four dimensional space-time, ($\theta=d\theta=0$) and write:
   \begin{equation}\label{lie2}
   \delta \mu^\mathcal{A}= \mu^\prime (x,o)-\mu^\mathcal{A}(x,0)=(\nabla\\epsilon)^\mathcal{A}+\epsilon^\alpha C_{\alpha|\mathcal{A}}R^\mathcal{A}_{ab},
   \end{equation}
   This is a supersymmetry transformation on space-time.\\

In order to obtain the space-time lagrangian we project the 4-form lagrangian from superspace
to space-time, namely we restrict all the terms to the
$\theta = 0\,,\,d \theta = 0$ hypersurface ${\cal M}^4$.
In practice one first goes
to the second order formalism by identifying the auxiliary 0-form
fields as explained before. Then one restricts the superfields to their
lowest ($\theta^\alpha = 0$) component and to the space-time bosonic vielbein or differentials. This gives the lagrangian 4-form (\ref{lag}). (For example the Lagrangian density in the usual tensor form is obtained by  expanding all the forms along the $dx^{\mu}$ differentials and taking the coefficient
of:
\begin{equation}
dx^{\mu}\wedge dx^{\nu}\wedge dx^{\rho}\wedge
dx^{\sigma}\,=\,{\epsilon^{\mu\nu\rho\sigma}\over \sqrt g}\left(
\sqrt g d^4x \right).
\end{equation}


\section{Normalizations and conventions}

\par
{\it Minkowski metric}:
\begin{equation}
\eta_{ab}\equiv \left(1,-1,-1,-1\right)
\end{equation}
 \par
{\it Definition of the Riemann tensor}:
\begin{equation}
R^{\mu}_{\phantom{\mu} \nu}\,=\, d \Gamma^{\mu}_{\phantom{\mu} \nu} +
\Gamma^{\mu}_{\phantom{\mu} \rho} \wedge \Gamma^{\rho}_{\phantom{\rho} \nu}
\equiv \,-\,{1 \over 2} R^{\mu}_{\phantom{\mu} \nu\rho \sigma}dx^\rho \wedge
dx^\sigma
\end{equation}

We are working with Majorana spinors, satisying: $ \bar \lambda = \lambda^T C$, where $C$ is the charge conjugation matrix.

 \textit{Symmetric $\gamma$ matrices: }
$\quad C \gamma_a\,,\quad C\gamma_{ab}$

\textit{Antisymmetric $\gamma$ matrices: }
$\quad C \,,\quad C\gamma_5\,,\quad C\gamma_5 \gamma_{a}$

{\it Clifford Algebra}:
\begin{eqnarray}
 \left\{\gamma_a,\gamma_b \right\}\, &=& \,2\,\eta_{ab} \nonumber\\
   \left[\gamma_a,\gamma_b \right]\,&=&\,2\,\gamma_{ab} \nonumber\\
  \gamma_5 \,& \equiv & -\, {\rm i}\, \gamma_0 \gamma_1 \gamma_2 \gamma_3
\nonumber\\
  \gamma _0^{\dagger}\,&=&\,\gamma _0; \qquad \qquad
 \gamma _0 \gamma _i^{\dagger}  \gamma _0 \,=\, \gamma _i  \qquad
 (i=1,2,3);\qquad \qquad  \gamma _5^{\dagger}\,=\,\gamma _5 \nonumber\\
  \epsilon _{abcd} \gamma^{cd}\,&=&\,2\,{\rm i}\, \gamma_{ab} \gamma_5\\
  \gamma_m \gamma^{ab}\gamma^m &=&0\\
\gamma_{ab}\gamma_m \gamma^{ab}&=&0\\
\gamma_{ab}\gamma_{cd} \gamma^{ab}&=&12 \gamma_{cd}\\
\gamma_m \gamma^{a }\gamma^m &=&-2\gamma^{a }\\
\gamma^{ab}\gamma^c &=&2 \gamma^{[a}\delta^{b]}_c + \ii \epsilon^{abcd}\gamma_5 \gamma_d\\
\gamma^c \gamma^{ab}&=& -2 \gamma^{[a}\delta^{b]}_c + \ii \epsilon^{abcd}\gamma_5 \gamma_d\\
\gamma_{ab}\gamma_{cd} &=&  \ii \epsilon^{abcd}\gamma_5 -4\delta^{[a}_{[c}\gamma^{b]}{\ d]} -2 \delta^{ab}_{cd}
 \label{clifdef}
\end{eqnarray}
\textit{Some useful Fierz identities for $N=1$} (for the 1-form spinor $\psi$):
\begin{eqnarray}
\psi \bar\psi&=& \frac 12 \gamma_a\bar\psi \gamma^a\psi-\frac 18 \gamma_{ab}\bar\psi\gamma^{ab}\psi\\
\gamma_a\psi\bar\psi \gamma^a\psi&=&0\\
\gamma_{ab}\psi\bar\psi\gamma^{ab}\psi&=&0\\
\gamma_{ab}\psi\bar\psi \gamma^a\psi&=&\psi\bar\psi \gamma_b\psi
\end{eqnarray}

\textit{Some useful relations and  Fierz identities for $N=2$} (for the 1-form spinors $\psi_A$, $\psi^A$):

$$\epsilon_{AB}= -\epsilon_{BA}\,,\quad \epsilon^{AB}= -\epsilon^{BA}\,,\quad \epsilon_{AB}\epsilon^{BC}= -\delta_A^C\,,$$
$$\epsilon_{AB}X^B = X_A\,,\quad \epsilon^{AB}X_B = -X^A$$
\begin{eqnarray}
\psi_A \bar\psi^B&=& \frac 12 \gamma_a\bar\psi^B \gamma^a\psi_A\\
\psi_A \bar\psi_B&=& \frac 12 \bar\psi_B \psi_A-\frac 18 \gamma_{ab}\bar\psi_B\gamma^{ab}\psi_A\\
\gamma_a\psi_A\bar\psi^B \gamma^a\psi_C&=&0\\
\gamma_{ab}\psi_A\bar\psi_B\gamma^{ab}\psi_C&=&4 \epsilon_{A(B}\psi_{C)}\bar\psi_L\psi_M\epsilon^{LM}\\
\gamma_{ab}\psi^A\bar\psi_B\gamma^{ab}\psi_C&=&0
\end{eqnarray}

{\it Reality condition for $SU(2)$ valued matrices}:
\begin{equation}
\bar S^{AB} \,=\,\epsilon ^{AC}\,\epsilon^{BD} (S_{CD})^{\star}
\end{equation}


\begin{thebibliography}{99}


\bibitem{York:1972sj}
  J.~W.~York, Jr.,
  ``Role of conformal three geometry in the dynamics of gravitation,''
  Phys.\ Rev.\ Lett.\  {\bf 28} (1972) 1082.
 \\
  J.~D.~Brown and J.~W.~York, Jr.,
  ``Quasilocal energy and conserved charges derived from the gravitational action,''
  Phys.\ Rev.\ D {\bf 47} (1993) 1407
  [gr-qc/9209012].

\bibitem{Gibbons:1976ue}
  G.~W.~Gibbons and S.~W.~Hawking,
  ``Action Integrals and Partition Functions in Quantum Gravity,''
  Phys.\ Rev.\ D {\bf 15} (1977) 2752.

\bibitem{Horava:1996ma}
  P.~Horava and E.~Witten,
  ``Eleven-dimensional supergravity on a manifold with boundary,''
  Nucl.\ Phys.\ B {\bf 475} (1996) 94
  [hep-th/9603142].

\bibitem{Maldacena:1997re}
  J.~M.~Maldacena,
  ``The Large N limit of superconformal field theories and supergravity,''
  Adv.\ Theor.\ Math.\ Phys.\  {\bf 2} (1998) 231
  [hep-th/9711200].
\\
  S.~S.~Gubser, I.~R.~Klebanov and A.~M.~Polyakov,
  ``Gauge theory correlators from noncritical string theory,''
  Phys.\ Lett.\ B {\bf 428} (1998) 105
  [hep-th/9802109].
\\
  E.~Witten,
  ``Anti-de Sitter space and holography,''
  Adv.\ Theor.\ Math.\ Phys.\  {\bf 2} (1998) 253
  [hep-th/9802150].
\\
  O.~Aharony, S.~S.~Gubser, J.~M.~Maldacena, H.~Ooguri and Y.~Oz,
  ``Large N field theories, string theory and gravity,''
  Phys.\ Rept.\  {\bf 323} (2000) 183
  [hep-th/9905111].

\bibitem{holren}
  V.~Balasubramanian and P.~Kraus,
  ``A Stress tensor for Anti-de Sitter gravity,''
  Commun.\ Math.\ Phys.\  {\bf 208} (1999) 413
  [hep-th/9902121].
\\
  J.~de Boer, E.~P.~Verlinde and H.~L.~Verlinde,
  ``On the holographic renormalization group,''
  JHEP {\bf 0008} (2000) 003
  [hep-th/9912012].
\\
  E.~P.~Verlinde and H.~L.~Verlinde,
  ``RG flow, gravity and the cosmological constant,''
  JHEP {\bf 0005} (2000) 034
  [hep-th/9912018].
\\
  J.~de Boer,
  ``The Holographic renormalization group,''
  Fortsch.\ Phys.\  {\bf 49} (2001) 339
  [hep-th/0101026].
 \\
  S.~de Haro, S.~N.~Solodukhin and K.~Skenderis,
  ``Holographic reconstruction of space-time and renormalization in the AdS / CFT correspondence,''
  Commun.\ Math.\ Phys.\  {\bf 217} (2001) 595
  [hep-th/0002230].
  \\
  K.~Skenderis,
  ``Lecture notes on holographic renormalization,''
  Class.\ Quant.\ Grav.\  {\bf 19} (2002) 5849
  [hep-th/0209067].
 \\
For more recent results, see also:
  A.~Lawrence and A.~Sever,
  ``Holography and renormalization in Lorentzian signature,''
  JHEP {\bf 0610} (2006) 013
  [hep-th/0606022].
\\
  I.~Heemskerk and J.~Polchinski,
  ``Holographic and Wilsonian Renormalization Groups,''
  JHEP {\bf 1106} (2011) 031
  [arXiv:1010.1264 [hep-th]].

\bibitem{Aros:1999id}
  R.~Aros, M.~Contreras, R.~Olea, R.~Troncoso and J.~Zanelli,
  ``Conserved charges for gravity with locally AdS asymptotics,''
  Phys.\ Rev.\ Lett.\  {\bf 84} (2000) 1647
  [gr-qc/9909015].
\\
  R.~Aros, M.~Contreras, R.~Olea, R.~Troncoso and J.~Zanelli,
  ``Conserved charges for even dimensional asymptotically AdS gravity theories,''
  Phys.\ Rev.\ D {\bf 62} (2000) 044002
  [hep-th/9912045].
\\
  P.~Mora, R.~Olea, R.~Troncoso and J.~Zanelli,
  ``Finite action principle for Chern-Simons AdS gravity,''
  JHEP {\bf 0406} (2004) 036
  [hep-th/0405267].
\\
  R.~Olea,
  ``Mass, angular momentum and thermodynamics in four-dimensional Kerr-AdS black holes,''
  JHEP {\bf 0506} (2005) 023
  [hep-th/0504233].
\\
  D.~P.~Jatkar, G.~Kofinas, O.~Miskovic and R.~Olea,
  ``Conformal Mass in AdS gravity,''
  arXiv:1404.1411 [hep-th].




\bibitem{MacDowell:1977jt}
  S.~W.~MacDowell and F.~Mansouri,
  ``Unified Geometric Theory of Gravity and Supergravity,''
  Phys.\ Rev.\ Lett.\  {\bf 38} (1977) 739
   [Erratum-ibid.\  {\bf 38} (1977) 1376].

   \bibitem{Andrianopoli:1996cm}
  L.~Andrianopoli, M.~Bertolini, A.~Ceresole, R.~D'Auria, S.~Ferrara, P.~Fre and T.~Magri,
  ``N=2 supergravity and N=2 superYang-Mills theory on general scalar manifolds: Symplectic covariance, gaugings and the momentum map,''
  J.\ Geom.\ Phys.\  {\bf 23} (1997) 111
  [hep-th/9605032].









\bibitem{vanNieuwenhuizen:2005kg}
  P.~van Nieuwenhuizen and D.~V.~Vassilevich,
  ``Consistent boundary conditions for supergravity,''
  Class.\ Quant.\ Grav.\  {\bf 22} (2005) 5029
  [hep-th/0507172].
\\
  D.~V.~Belyaev,
  ``Boundary conditions in supergravity on a manifold with boundary,''
  JHEP {\bf 0601} (2006) 047
  [hep-th/0509172].
\\
  D.~V.~Belyaev and P.~van Nieuwenhuizen,
  ``Tensor calculus for supergravity on a manifold with boundary,''
  JHEP {\bf 0802} (2008) 047
  [arXiv:0711.2272 [hep-th]].
\\
  D.~V.~Belyaev and P.~van Nieuwenhuizen,
  ``Simple d=4 supergravity with a boundary,''
  JHEP {\bf 0809} (2008) 069
  [arXiv:0806.4723 [hep-th]].
\\
  D.~Grumiller and P.~van Nieuwenhuizen,
  ``Holographic counterterms from local supersymmetry without boundary conditions,''
  Phys.\ Lett.\ B {\bf 682} (2010) 462
  [arXiv:0908.3486 [hep-th]].
  D.~V.~Belyaev and T.~G.~Pugh,
  ``The Supermultiplet of boundary conditions in supergravity,''
  JHEP {\bf 1010} (2010) 031
  [arXiv:1008.1574 [hep-th]].

\bibitem{esposito}
  G.~Esposito, A.~Y.~.Kamenshchik and K.~Kirsten,
  ``One loop effective action for Euclidean Maxwell theory on manifolds with boundary,''
  Phys.\ Rev.\ D {\bf 54} (1996) 7328
  [hep-th/9606132].
  \\
  I.~G.~Avramidi and G.~Esposito,
  ``Gauge theories on manifolds with boundary,''
  Commun.\ Math.\ Phys.\  {\bf 200} (1999) 495
  [hep-th/9710048].

\bibitem{Moss:2003bk}
  I.~G.~Moss,
  ``Boundary terms for eleven-dimensional supergravity and M theory,''
  Phys.\ Lett.\ B {\bf 577} (2003) 71
  [hep-th/0308159].
\\
  I.~GMoss,
  ``Boundary terms for supergravity and heterotic M theory,''
  Nucl.\ Phys.\ B {\bf 729} (2005) 179
  [hep-th/0403106].

\bibitem{Howe:2011tm}
  P.~S.~Howe, T.~G.~Pugh, K.~S.~Stelle and C.~Strickland-Constable,
  ``Ectoplasm with an Edge,''
  JHEP {\bf 1108} (2011) 081
  [arXiv:1104.4387 [hep-th]].

\bibitem{librone}
An extended dissertation on the geometric approach to supergravity can be found in L.~Castellani, R.~D'Auria and P.~Fre,
  ``Supergravity and superstrings: A Geometric perspective. Vol. 2: Supergravity,''
  Singapore, Singapore: World Scientific (1991) 607-1371. A shorter account of the approach is also given in the Appendices of \cite{Andrianopoli:1996cm}.


\end{thebibliography}
\end{document}